\documentclass[11pt, a4paper, oneside]{article}
\usepackage{amsmath}
\usepackage{amssymb}
\usepackage{amsfonts}

\input{tcilatex}
\begin{document}

\newpage 
\begin{titlepage}
\begin{flushright}
UFSCARF-TH-10-08
\end{flushright}
\vskip 3.0cm
\begin{center}
{\Large  Reflection matrices for the $U_{q}[spo(2n|2m)]$ vertex model }\\
\vskip 2cm
{\large A. Lima-Santos} \\
\vskip 1cm
{\em Universidade Federal de S\~ao Carlos\\
Departamento de F\'{\i}sica \\
C.P. 676, 13565-905, S\~ao Carlos-SP, Brasil}\\
\end{center}
\vskip 2cm

\begin{abstract}
We propose a classification of  the solutions of the graded reflection equations to the $U_{q}[spo(2n|2m)]$ 
vertex model. We find  twelve distinct classes of reflection matrices such that four of them are diagonal. 
In the non-diagonal matrices the number of free parameters depending on the number of bosonic ($2n$) and 
fermionic ($2m$) degrees of freedom while in the diagonal ones we find solutions with at most one free parameter.
\end{abstract}

\vskip 2.5cm
\centerline{{\small PACS numbers:  05.50+q, 02.30.IK, 75.10.Jm}}
\vskip 0.1cm
\centerline{{\small Keywords: Reflection Equations, K-matrices, Superalgebras}}
\vskip 2.5cm
\centerline{{\today}}
\end{titlepage}

\section{Introduction}

The present work is the third of three papers devoted to the classification
of the integrable reflection K-matrices for the vertex models associated
with the $U_{q}[sl(r|2m)^{(2)}]$, $U_{q}[osp(r|2m)^{(1)}]$ and $%
U_{q}[spo(2n|2m)]$ Lie superalgebras. In the first paper \cite{LIMG} we have
considered the $U_{q}[sl(r|2m)^{(2)}]$ vertex model and in the second paper 
\cite{LIMS}\ the $U_{q}[osp(r|2m)^{(1)}$ vertex model ones. In this paper we
have presented the general set of regular solutions of the graded reflection
equation for the $U_{q}[spo(2n|2m)]$\ vertex model.

It is well-known that the Lie superalgebra $osp(2m|2n)$ and $spo(2n|2m)$ are
naturally isomorphic: From the classical Lie superalgebras\cite{Kac}, we
have the notation $D(m|n)=osp(2m|2n)=spo(2n|2m)$. Thus, the transition from $%
osp(2m|2n)^{(1)}$ to $spo(2n|2m)$ amounts to shift of the gradation of the
vector representation\cite{Scheunert}. But, we will work with the natural
gradations in order to avoid any shift of gradations. Moreover, for trivial
values of $m$ and $n$, a Lie superalgebra coincides with a Lie algebra: $%
D(m|0)=D_{m}^{(1)}$, $D(0|n)=C_{n}^{(1)}.$Therefore, by considering the $%
U_{q}[spo(2n|2m)]$ vertex model we are completing the study of the
integrable $K$-matrices for the graded version of the vertex models
associated with the $B_{n}^{(1)},C_{n}^{(1)},D_{n}^{(1)},A_{2n}^{(2)}$ and $%
A_{2n-1}^{(2)}$ Lie algebras.

Our findings can be summarized into four classes of diagonal solutions and
eight classes of non-diagonal ones. This paper is organized as follows. In
the next section we present the $R$-matrix of the $U_{q}[spo(2n|2m)]$ vertex
model in terms of standard Weyl matrices. In the section $3$ we present four
classes of diagonal solutions. In the section $4$ we present eight classes
of non-diagonal solutions what we hope to be the most general set of $K$%
-matrices for the vertex model here considered. Concluding remarks are
discussed in the section $5$, and in the appendix $A$ we present solutions
associated with the $U_{q}[spo(2|2)]$ case.

\section{The $U_{q}[spo(2n|2m)]$ vertex model}

\noindent The $U_{q}[spo(2n|2m)]$ invariant $R$-matrices are given by\cite%
{GALLEAS} 
\begin{eqnarray}
R(x) &=&\sum_{\overset{i=1}{i\neq i^{\prime }}}^{N}(-1)^{p_{i}}a_{i}(x)\hat{e%
}_{ii}\otimes \hat{e}_{ii}+b(x)\sum_{\overset{i,j=1}{i\neq j,i\neq j^{\prime
}}}^{N}\hat{e}_{ii}\otimes \hat{e}_{jj}  \notag \\
&&+{\bar{c}}(x)\sum_{\overset{i,j=1}{i<j,i\neq j^{\prime }}}(-1)^{p_{i}p_{j}}%
\hat{e}_{ji}\otimes \hat{e}_{ij}+c(x)\sum_{\overset{i,j=1}{i>j,i\neq
j^{\prime }}}^{N}(-1)^{p_{i}p_{j}}\hat{e}_{ji}\otimes \hat{e}_{ij}  \notag \\
&&+\sum_{i,j=1}^{N}(-1)^{p_{i}}d_{i,j}(x)\hat{e}_{ij}\otimes \hat{e}%
_{i^{\prime }j^{\prime }}  \label{Rspo}
\end{eqnarray}%
where $N=2n+2m$ is the dimension of the graded space with $2n$ bosonic and $%
2m$ fermionic degrees of freedom. Here $i^{\prime }=N+1-i$ corresponds to
the conjugated index of $i$ and $\hat{e}_{ij}$ refers to a usual $N\times N$
Weyl matrix with only one non-null entry with value $1$ at the row $i$ and
column $j$.

In what follows we shall adopt the grading structure 
\begin{equation}
p_{i}=\left\{ 
\begin{array}{c}
0\qquad \quad \mathrm{for\ }i=1,...,m\ \mathrm{\ and\ \ }i=2n+m+1,...,N \\ 
\ 1\qquad \mathrm{for\ }i=m+1,...,2n+m\qquad \qquad \mathrm{\qquad \qquad
\quad }%
\end{array}%
\right. ,  \label{grad}
\end{equation}%
and the corresponding Boltzmann weights $a_{i}(x)$, $b(x)$, $c(x)$, $\bar{c}%
(x)$ and $d_{ij}(x)$ are then given by%
\begin{eqnarray}
a_{i}(x) &=&(x-\zeta )(x^{(1-p_{i})}-q^{2}x^{p_{i}}),\quad \
b(x)=q(x-1)(x-\zeta ),  \notag \\
c(x) &=&(1-q^{2})(x-\zeta ),\qquad \qquad \quad {\bar{c}}(x)=x(1-q^{2})(x-%
\zeta )
\end{eqnarray}%
and 
\begin{equation}
\;d_{i,j}(x)=\left\{ 
\begin{array}{c}
q(x-1)(x-\zeta )+x(q^{2}-1)(\zeta -1),\qquad \quad \quad 
\hfill%
(i=j=j^{\prime }) \\ 
(x-1)[(x-\zeta )(-1)^{p_{i}}q^{2p_{i}}+x(q^{2}-1)],\qquad 
\hfill%
(i=j\neq j^{\prime }) \\ 
(q^{2}-1)[\zeta (x-1)\frac{\theta _{i}q^{t_{i}}}{\theta _{j}q^{t_{j}}}%
-\delta _{i,j^{\prime }}(x-\zeta )],\qquad \quad \quad 
\hfill%
(i<j) \\ 
(q^{2}-1)x[(x-1)\frac{\theta _{i}q^{t_{i}}}{\theta _{j}q^{t_{j}}}-\delta
_{i,j^{\prime }}(x-\zeta )],\qquad \quad \quad 
\hfill%
(i>j)%
\end{array}%
\right.
\end{equation}%
where $\zeta =q^{2n-2m+2}$. The remaining variables $\theta _{i}$ and $t_{i}$
depend strongly on the grading structure considered and they are determined
by the relations%
\begin{equation}
\theta _{i}=\left\{ 
\begin{array}{c}
(-1)^{-\frac{p_{i}}{2}},\qquad 
\hfill%
1\leq i\leq \frac{N}{2} \\ 
-(-1)^{\frac{p_{i}}{2}},\qquad 
\hfill%
\frac{N}{2}+1\leq i\leq N%
\end{array}%
\right.
\end{equation}%
\begin{equation}
t_{i}=\left\{ 
\begin{array}{c}
\displaystyle%
i-[\frac{1}{2}+p_{i}-2\sum_{j=i}^{\frac{N}{2}}p_{j}],\qquad 
\hfill%
1\leq i\leq \frac{N}{2} \\ 
\\ 
\displaystyle%
i+[\frac{1}{2}+p_{i}-2\sum_{j=\frac{N}{2}+1}^{i}p_{j}],\qquad 
\hfill%
\frac{N}{2}+1\leq i\leq N%
\end{array}%
\right.
\end{equation}

The $R$-matrix (\ref{Rspo}) satisfies important symmetry relations, besides
the standard properties of regularity and unitarity, namely 
\begin{eqnarray}
&&\mathrm{PT-Symmetry:\;}\;\;\;\;\;\;\;\;\;\;\;\;\;\;%
\;R_{21}(x)=R_{12}^{st_{1}st_{2}}(x)  \notag \\
&&\mathrm{Cros}\text{\textrm{sing}}\mathrm{\;Symmetry:}\;\;\;\;\;\;R_{12}(x)=%
\frac{\rho (x)}{\rho (x^{-1}\eta ^{-1})}V_{1}R_{12}^{st_{2}}(x^{-1}\eta
^{-1})V_{1}^{-1},
\end{eqnarray}%
where the symbol $st_{k}$ stands for the supertransposition operation in the
space with index $k$. In its turn $\rho (x)$ is an appropriate normalization
function given by $\rho (x)=q(x-1)(x-\zeta )$ and the crossing parameter is $%
\eta =\zeta ^{-1}$. The corresponding crossing matrix $V$ is an
anti-diagonal matrix with the following non-null entries $V_{i{i}^{^{\prime
}}}$ 
\begin{equation}
V_{i{i}^{^{\prime }}}=.\left\{ 
\begin{array}{c}
\displaystyle%
(-1)^{\frac{1-p_{i}}{2}},\qquad 
\hfill%
i=1, \\ 
\displaystyle%
(-1)^{\frac{1-p_{i}}{2}}q^{\digamma _{1}^{(i)}},\qquad 
\hfill%
1<i<\frac{N+1}{2}, \\ 
\displaystyle%
(-1)^{\frac{1+p_{i}}{2}}q^{\digamma _{2}^{(i)}},\qquad 
\hfill%
\frac{N+1}{2}<i\leq N.%
\end{array}%
\right.
\end{equation}%
where%
\begin{equation}
\displaystyle%
\digamma _{1}^{(i)}=i-1-p_{1}-p_{i}-2\sum_{j=1}^{i-1}p_{j},\quad 
\displaystyle%
\digamma _{2}^{(i)}=i-p_{1}-p_{i}-2\sum_{j=2\neq \frac{N}{2}+1}^{i-1}p_{j}.
\label{bwf}
\end{equation}

The construction of integrable models with open boundaries was largely
impulsed by Sklyanin's pioneer work \cite{SK}. In Sklyanin's approach the
construction of such models are based on solutions of the so-called
reflection equations \cite{CHER} for a given integrable bulk system. The
reflection equations determine the boundary conditions compatible with the
bulk integrability and it reads 
\begin{equation}
R_{21}(x/y)K_{2}^{-}(x)R_{12}(xy)K_{1}^{-}(y)=K_{1}^{-}(y)R_{21}(xy)K_{2}^{-}(x)R_{12}(x/y),
\label{RE}
\end{equation}%
where the tensor products appearing in (\ref{RE}) should be understood in
the graded sense. The matrix $K^{-}(x)$ describes the reflection at one of
the ends of an open chain while a similar equation should also hold for a
matrix $K^{+}(x)$ describing the reflection at the opposite boundary. As
discussed above, the $U_{q}[spo(2n|2m)]$ $R$-matrix satisfies important
symmetry relations such as the PT-symmetry and crossing symmetry. When these
properties are fulfilled one can follow the scheme devised in \cite{BRA,MEZ}
and the matrix $K^{-}(x)$ is obtained by solving the Eq. (\ref{RE}) while
the matrix $K^{+}(x)$ can be obtained from the isomorphism $K^{-}(x)\mapsto
K^{+}(x)^{st}=K^{-}(x^{-1}\eta ^{-1})V^{st}V$.

The purpose of this work is to investigate the general families of regular
solutions of the graded reflection equation (\ref{RE}). Regular solutions
mean that the $K$-matrices have the general form 
\begin{equation}
K^{-}(x)=\sum_{i,j=1}^{N}k_{i,j}(x)\;\hat{e}_{ij},  \label{KM}
\end{equation}%
such that the condition $k_{i,j}(1)=\delta _{ij}$ holds for all matrix
elements.

The direct substitution of (\ref{KM}) and the $U_{q}[spo(2n|2m)]$ $R$-matrix
(\ref{Rspo})-(\ref{bwf}) in the graded reflection equation (\ref{RE}), leave
us with a system of $N^{4}$ functional equations for the entries $k_{i,j}(x)$%
. In order to solve these equations we shall make use of the derivative
method. Thus, by differentiating the equation (\ref{RE}) with respect to $y$
and setting $y=1$, we obtain a set of algebraic equations for the matrix
elements $k_{i,j}(x)$. Although we obtain a large number of equations only a
few of them are actually independent and a direct inspection of those
equations, in the lines described in \cite{LIM}, allows us to find the
branches of regular solutions. In what follows we shall present our findings
for the regular solutions of the reflection equation associated with the $%
U_{q}[spo(2n|2m)]$ vertex model.

\section{Diagonal K-matrix solutions}

The diagonal solutions of the graded reflection equation (\ref{RE}) is
characterized by a $K$-matrix of the form 
\begin{equation}
K^{-}(x)=\sum_{i=1}^{N}k_{i,i}(x)\hat{e}_{ii}.
\end{equation}%
with the entries $k_{i,j}(x)$ related with $k_{1,1}(x)$ in a general form
given by%
\begin{equation}
k_{i,i}(x)=\frac{(\beta _{i,i}-\beta _{1,1})(x-1)+2}{(\beta _{i,i}-\beta
_{1,1})(x^{-1}-1)+2}k_{1,1}(x)
\end{equation}%
for $i=2,...,N-1$ and%
\begin{equation}
k_{N,N}(x)=\frac{(\beta _{N,N}-\beta _{1,1})(x-1)+2}{(\beta _{N,N}-\beta
_{1,1})(x^{-1}-1)+2}k_{N-1,N-1}(x)
\end{equation}%
The parameters $\beta _{i,i}=\frac{d}{dx}[k_{i,i}(x)]_{x=1}$ are fixed in
order to give us four families of diagonal $K$-matrices such that half of
them with one free parameter $\beta $.%
\begin{eqnarray}
k_{1,1}(x) &=&%
\displaystyle%
\frac{\beta (x^{-1}-1)+2}{\beta (x-1)+2},  \notag \\
k_{22}(x) &=&\cdots =k_{N-1,N-1}(x)=1,  \notag \\
k_{N,N}(x) &=&%
\displaystyle%
x\frac{\beta (1-xq^{2}\zeta )-2}{\beta (x-q^{2}\zeta )-2x}.  \label{d.1}
\end{eqnarray}%
and%
\begin{eqnarray}
k_{1,1}(x) &=&\cdots =k_{n+m,n+m}(x)=1,  \notag \\
\displaystyle%
k_{n+m+1,m+m+1}(x) &=&\cdots =k_{N,N}(x)=\frac{\beta (x-1)+2}{\beta
(x^{-1}-1)+2}.  \label{d.2}
\end{eqnarray}%
while the other half without free parameters: 
\begin{eqnarray}
k_{1,1}(x) &=&\cdots =k_{p-1,p-1}(x)=1,  \notag \\
\displaystyle%
k_{p,p}(x) &=&\cdots =k_{N+1-p,N+1-p}(x)=\frac{x+\epsilon q^{2p-3}\sqrt{%
\zeta }}{x^{-1}+\epsilon q^{2p-3}\sqrt{\zeta }},  \notag \\
k_{N+2-p,N+2-p}(x) &=&\cdots =k_{N,N}(x)=x^{2}.  \label{d.3}
\end{eqnarray}%
where the discrete label $p$ assumes values in the interval $2\leq p\leq m+1$
\ and 
\begin{eqnarray}
k_{1,1}(x) &=&\cdots =k_{p-1,p-1}(x)=1,  \notag \\
k_{p,p}(x) &=&\cdots =k_{N+1-p,N+1-p}(x)=%
\displaystyle%
\frac{x+\epsilon q^{4m+1-2p}\sqrt{\zeta }}{x^{-1}+\epsilon q^{4m+1-2p}\sqrt{%
\zeta }},  \notag \\
k_{N+2-p,N+2-p}(x) &=&\cdots =k_{N,N}(x)=x^{2}.  \label{d.4}
\end{eqnarray}%
for the discrete label $p$ with values in the interval $m+2\leq p\leq n+m.$

Here and in what follows, $\epsilon $ is a discrete parameter assuming the
values $\pm 1$. We also notice that the case\ $p=2$ is limit of the solution
of the solution (\ref{d.1}) and we have $(n+m)$ diagonal solutions.

\section{Non-Diagonal K-matrix Solutions}

\bigskip Analyzing the reflection matrix equation (\ref{RE}) we can see that
the non-diagonal matrix elements $k_{i,j}(x)$ have the form 
\begin{equation}
k_{i,j}(x)=\left\{ 
\begin{array}{c}
\beta _{i,j}xG(x),\quad 
\hfill%
\mathrm{\ }i>j^{\prime } \\ 
\beta _{i,j}xG(x),\quad 
\hfill%
i>j^{\prime } \\ 
\beta _{i,j}G(x)H_{f}(x),\quad 
\hfill%
\mathrm{\quad }i=j^{\prime }(\ \mathrm{fermionic}) \\ 
\beta _{i,j}G(x)H_{b}(x),\quad 
\hfill%
\mathrm{\quad }i=j^{\prime }(\ \mathrm{bosonic})%
\end{array}%
\right.   \label{Gf}
\end{equation}%
Here and in what follows $G(x)$ is an arbitrary function satisfying the
regular condition $k_{i,j}(1)=\delta _{ij}$ and%
\begin{equation}
\beta _{i,j}=\frac{d}{dx}[k_{i,j}(x)]_{x=1},\qquad H_{f}(x)=\frac{x-\epsilon
q\sqrt{\zeta }}{1-\epsilon q\sqrt{\zeta }},\qquad H_{b}(x)=\frac{qx+\epsilon 
\sqrt{\zeta }}{q+\epsilon \sqrt{\zeta }}.  \label{Hs}
\end{equation}%
Moreover, the diagonal entries $k_{i,i}(x)$ satisfy defined recurrence
relations which depend on the bosonic and fermionic degree of freedom.
However, for the $U_{q}[spo(2n|2m)]$ \ model we have found $K$-matrix
solutions with both degree of freedom only for the cases $(n\neq 1,m=1)$ and 
$(n=1,m\neq 1)$.

\subsection{K-matrices with fermionic degree of freedom}

Here we shall focus on the non-diagonal solutions of the graded reflection
equation (\ref{RE})\ with $k_{i,j}(x)=0$ for the bosonic degree of freedom \ 
\textit{i.e.}, \ for $i\neq j=m+1,...,m+2n$. We have found three classes of
non-diagonal solutions that we refer in what follows as solutions of type $%
\mathcal{P}_{1}$ to type $\mathcal{P}_{3}$:

\subsubsection{\textbf{Solution $\mathcal{P}_{1}$}}

The solution of type $\mathcal{P}_{1}$ is valid only for the $%
U_{q}[sop(2n|2)]$ models with $n\geq 1$ and the $K$-matrix has the following
block structure 
\begin{equation}
K^{-}=\left( 
\begin{array}{ccc}
k_{1,1} & \mathbb{O}_{1\times 2n} & k_{1,N} \\ 
\mathbb{O}_{2n\times 1} & \mathbb{K}_{1} & \mathbb{O}_{2n\times 1} \\ 
k_{N,1} & \mathbb{O}_{1\times 2n} & k_{N,N}%
\end{array}%
\right) ,
\end{equation}%
where $\mathbb{O}_{a\times b}$ is a $a\times b$ null matrix and 
\begin{equation}
\mathbb{K}_{1}(x)=\frac{x^{2}-\zeta }{1-\zeta }\mathbb{I}_{2n\times 2n}.
\end{equation}%
Here and in what follows $\mathbb{I}_{2n\times 2n}$ denotes a $2n\times 2n$
identity matrix and the remaining non-null entries are given by 
\begin{eqnarray}
k_{1,1}(x) &=&1,  \notag \\
k_{1,N}(x) &=&\frac{1}{2}\beta _{1,N}(x^{2}-1),\quad k_{N,1}(x)=\frac{2\zeta 
}{(1-\zeta )^{2}}\frac{(x^{2}-1)}{\beta _{1,N}},\;  \notag \\
k_{N,N}(x) &=&x^{2}.
\end{eqnarray}%
where $\beta _{1,N}$ is the free parameter. We remark here that this
solution for $n=1$ consist of a particular case of the three parameter
solution given in the appendix for the $U_{q}[spo(2|2)]$ vertex model.

\subsubsection{\textbf{Solution $\mathcal{P}_{2}$}}

The $U_{q}[spo(2n|4)]$ vertex models admit the solution $\mathcal{P}_{2}$
whose corresponding $K$-matrix has the following structure 
\begin{equation}
K^{-}=\left( 
\begin{array}{ccccc}
k_{1,1} & k_{1,2} &  & k_{1,N-1} & k_{1,N} \\ 
k_{2,1} & k_{2,2} & \mathbb{O}_{2\times 2n} & k_{2,N-1} & k_{2,N} \\ 
& \mathbb{O}_{2n\times 2} & \mathbb{K}_{2} & \mathbb{O}_{2n\times 2} &  \\ 
k_{N-1,1} & k_{N-1,2} & \mathbb{O}_{2\times 2n} & k_{N-1,N-1} & k_{N-1,N} \\ 
k_{N,1} & k_{N,2} &  & k_{N,N-1} & k_{N,N}%
\end{array}%
\right) ,
\end{equation}%
where $\mathbb{K}_{2}=k_{3,3}(x)\mathbb{I}_{2n\times 2n}$. The non-diagonal
entries can be written as 
\begin{eqnarray}
k_{1,2}(x) &=&\beta
_{1,2}G_{1}(x),\;\;\;\;\;\;\;\;\;\;\;\;\;\;\;\;\;\;\;\;\;\;\;\;\;\;\;\;\;\;%
\;\;k_{2,1}(x)=\beta _{2,1}G_{1}(x)  \notag \\
k_{1,N-1}(x) &=&\beta
_{1,N-1}G_{1}(x)\;\;\;\;\;\;\;\;\;\;\;\;\;\;\;\;\;\;\;\;\;\;\;\;\;\;%
\;k_{N-1,1}(x)=q^{2}\zeta \frac{\beta _{1,2}\beta _{2,1}}{\beta _{1,N-1}}%
G_{1}(x)  \notag \\
k_{2,N-1}(x) &=&-\frac{\beta _{2,1}\beta _{1,N}}{\beta _{1,2}}%
G_{1}(x)H(x)\;\;\;\;\;\;\;\;\;\;\;\;\;\;k_{N-1,2}(x)=-q^{2}\zeta \frac{\beta
_{1,2}\beta _{2,1}\beta _{1,N}}{\beta _{1,N-1}^{2}}G_{1}(x)H(x)  \notag \\
k_{2,N}(x) &=&-\Gamma _{n}\frac{\beta _{1,N}}{\beta _{1,2}}%
xG_{2}(x)\;\;\;\;\;\;\;\;\;\;\;\;\;\;\;\;\;\;\;\;\;\;\;k_{N,2}(x)=-q^{2}%
\zeta \Gamma _{n}\frac{\beta _{2,1}\beta _{1,N}}{\beta _{1,N-1}^{2}}xG_{2}(x)
\notag \\
k_{N-1,N}(x) &=&-q^{2}\zeta \Gamma _{n}\frac{\beta _{1,N}}{\beta _{1,N-1}}%
xG_{2}(x)\;\;\;\;\;\;\;\;\;\;\;\;\;\;\;\;k_{N,N-1}(x)=-q^{2}\zeta \Gamma _{n}%
\frac{\beta _{2,1}\beta _{1,N}}{\beta _{1,2}\beta _{1,N-1}}xG_{2}(x)  \notag
\\
k_{N,1}(x) &=&q^{2}\zeta \frac{\beta _{2,1}^{2}\beta _{1,N}}{\beta
_{1,N-1}^{2}}G_{1}(x)H(x)\;\;\;\;\;\;\;\;\;\;\;\;\;k_{1,N}(x)=\beta
_{1,N}G_{1}(x)H(x),
\end{eqnarray}%
and the auxiliary functions are given by%
\begin{eqnarray}
G_{1}(x) &=&\left[ -q^{2}\zeta \Gamma _{n}\frac{\beta _{1,N}}{\beta
_{1,2}\beta _{1,N-1}}+\frac{x-q^{2}\zeta }{x-1}\right] \left( \frac{x-1}{%
x^{2}-q^{2}\zeta }\right) \frac{k_{1,N}(x)}{\beta _{1,N}},  \notag \\
G_{2}(x) &=&\left[ -\frac{1}{\Gamma _{n}}\frac{\beta _{1,2}\beta _{1,N-1}}{%
\beta _{1,N}}+\frac{x-q^{2}\zeta }{x-1}\right] \left( \frac{x-1}{%
x^{2}-q^{2}\zeta }\right) \frac{k_{1,N}(x)}{\beta _{1,N}},  \notag \\
H(x) &=&\frac{\beta _{1,2}\beta _{1,N-1}(x^{2}-q^{2}\zeta )}{\beta
_{1,2}\beta _{1,N-1}(x-q^{2}\zeta )-q^{2}\zeta \Gamma _{n}\beta _{1,N}(x-1)}
\end{eqnarray}%
and%
\begin{equation}
\Gamma _{n}=\frac{\beta _{2,1}\beta _{1,N}}{\beta _{1,N-1}}-\frac{2}{%
1-q^{2}\zeta }
\end{equation}

With respect to the diagonal matrix elements, we have the following
expressions 
\begin{eqnarray}
k_{11}(x) &=&\left\{ (x-q^{2}\zeta )\left[ (1+q^{2}\zeta )\frac{\beta
_{2,1}\beta _{1,N}}{\beta _{1,N-1}}-\frac{\beta _{1,2}\beta _{1,N-1}}{\beta
_{1,N}}\right] +q^{2}\zeta \Gamma _{n}^{2}(1-xq^{2}\zeta )\frac{\beta _{1,N}%
}{\beta _{1,2}\beta _{1,N-1}}\right.  \notag \\
&&+\left. \frac{2(1+q^{2}\zeta )^{2}x-4q^{2}\zeta (x^{2}+1)}{(1-q^{2}\zeta
)(x-1)}\right\} \frac{1}{(x^{2}-q^{2}\zeta )(x+1)}\frac{k_{1,N}(x)}{\beta
_{1,N}}
\end{eqnarray}%
for the recurrence relation%
\begin{eqnarray}
k_{2,2}(x) &=&k_{1,1}(x)+(\beta _{2,2}-\beta _{1,1})G_{1}(x),  \notag \\
k_{3,3}(x) &=&k_{1,1}(x)+(\beta _{3,3}-\beta _{1,1})G(x)+\Delta _{1}(x), 
\notag \\
k_{N-1,N-1}(x) &=&k_{3,3}(x)+\left( \beta _{N-1,N-1}-\beta _{3,3}\right)
xG_{2}(x)+\Delta _{2}(x),  \notag \\
k_{N,N}(x) &=&k_{N-1,N-1}(x)+\left( \beta _{N,N}-\beta _{N-1,N-1}\right)
xG_{2}(x).  \label{dd}
\end{eqnarray}%
where 
\begin{eqnarray}
\Delta _{1}(x) &=&\frac{\beta _{2,1}}{\beta _{1,N-1}}\left( x+q^{2}\zeta
\Gamma _{n}\frac{\beta _{1,N}}{\beta _{1,2}\beta _{1,N-1}}\right) \left( 
\frac{x-1}{x^{2}-q^{2}\zeta }\right) k_{1,N}(x)  \notag \\
\Delta _{2}(x) &=&-\frac{q^{2}\zeta }{\Gamma _{n}}\beta _{1,2}\frac{\beta
_{1,N-1}}{\beta _{1,N}}\Delta _{1}(x).
\end{eqnarray}%
The diagonal entries (\ref{dd}) depend on the variables $\beta _{\alpha
,\alpha }$ which are related to the\ four free parameters $\beta
_{1,2},\beta _{2,1},\beta _{1,N-1}$ and $\beta _{1,N}$ through the
expressions 
\begin{eqnarray}
\beta _{2,2} &=&\beta _{1,1}+\frac{\beta _{1,2}\beta _{1,N-1}}{\beta _{1,N}}%
-\Gamma _{n},  \notag \\
\beta _{3,3} &=&\beta _{1,1}+\frac{\beta _{1,2}\beta _{1,N-1}}{\beta _{1,N}}+%
\frac{2}{1-q^{2}\zeta },  \notag \\
\beta _{N-1,N-1} &=&\beta _{1,1}+2+\frac{\beta _{1,2}\beta _{1,N-1}}{\beta
_{1,N}}-q^{2}\zeta \Gamma _{n},  \notag \\
\beta _{N,N} &=&\beta _{1,1}+2+\frac{\beta _{1,2}\beta _{1,N-1}}{\beta _{1,N}%
}-q^{2}\zeta \Gamma _{n}^{2}\frac{\beta _{1,N}}{\beta _{1,2}\beta _{1,N-1}}.
\end{eqnarray}

\subsubsection{\textbf{Solution $\mathcal{P}_{3}$}}

This class of solution is valid for all $U_{q}[spo(2n|2m)]$ vertex models
with $m\geq 3$ and the corresponding $K$-matrix possess the following
general form 
\begin{equation}
K^{-}=\!\!\!\left( 
\begin{array}{ccccccc}
k_{1,1} & \cdots  & k_{1,m} & \cdots  & k_{1,2n+m+1} & \cdots  & k_{1,N} \\ 
\vdots  & \ddots  & \vdots  & \mathbb{O}_{m\times 2n} & \vdots  & \ddots  & 
\vdots  \\ 
k_{m,1} & \cdots  & k_{m,m} & \cdots  & k_{m,2n+m+1} & \cdots  & k_{m,N} \\ 
& \!\mathbb{O}_{2n\times m} &  & \!\mathbb{K}_{3} &  & \!\mathbb{O}%
_{2n\times m} &  \\ 
k_{2n+m+1,1} & \cdots  & k_{2n+m+1,m} & \cdots  & k_{2n+m+1,2n+m+1} & \cdots 
& k_{2n+m+1,N} \\ 
\vdots  & \ddots  & \vdots  & \mathbb{O}_{m\times 2n} & \vdots  & \ddots  & 
\vdots  \\ 
k_{N,1} & \cdots  & k_{N,m} & \cdots  & k_{N,2n+m+1} & \cdots  & k_{N,N}%
\end{array}%
\right) ,
\end{equation}%
where $\mathbb{K}_{3}$ is a diagonal matrix given by 
\begin{equation}
\mathbb{K}_{3}(x)=k_{m+1,m+1}(x)\;\mathbb{I}_{2n\times 2n}.
\end{equation}

With respect to the elements of the last column, we have the following
expression 
\begin{eqnarray}
k_{i,N}(x) &=&-\frac{\epsilon }{\sqrt{\zeta }}q^{t_{i}-t_{1}}\beta
_{1,i^{\prime }}xG(x) \\
\;i &=&2,\dots ,m\;\;\mathrm{and\quad }i=2n+m+1,\dots ,N-1,  \notag
\end{eqnarray}%
In their turn the entries of the first column are mainly given by 
\begin{eqnarray}
k_{i,1}(x) &=&q^{t_{i}-t_{2}}\frac{\beta _{2,1}\beta _{1,i^{\prime }}}{\beta
_{1,N-1}}G(x), \\
i &=&3,\dots ,m\;\;\mathrm{and\quad }i=2n+m+1,\dots ,N-1.  \notag
\end{eqnarray}%
In the last row we have 
\begin{eqnarray}
k_{N,j}(x) &=&-\frac{\epsilon }{\sqrt{\zeta }}q^{t_{N}-t_{2}}\frac{\beta
_{2,1}\beta _{1,j}}{\beta _{1,N-1}}xG(x) \\
j &=&2,\dots ,m\;\;\mathrm{and\quad }j=2n+m+1,\dots ,N-1,  \notag
\end{eqnarray}%
while the elements of the first row are $k_{1,j}(x)=\beta _{1,j}G(x)$ for $%
j=2,\dots ,m$ and\ $j=2n+m+1,\dots ,N-1$.

Concerning the elements of the secondary diagonal, they are given by 
\begin{eqnarray}
k_{i,i^{\prime }}(x) &=&q^{t_{1}-t_{i^{\prime }}}\left( \frac{1-\epsilon q%
\sqrt{\zeta }}{q+1}\right) ^{2}\frac{\beta _{1,i^{\prime }}^{2}}{\beta _{1,N}%
}G(x)H_{f}(x)  \notag \\
i &=&2,\dots ,m\;,\;i\neq i^{\prime }\quad \mathrm{and\quad }i=2n+m+1,\dots
,N-1,
\end{eqnarray}%
while the remaining entries $k_{1,N}(x)$ and $k_{N,1}(x)$ are determined by
the following expressions 
\begin{eqnarray}
k_{1,N}(x) &=&\beta _{1,N}G(x)H_{f}(x)  \notag \\
k_{N,1}(x) &=&q^{t_{N-1}-t_{2}}\frac{\beta _{1,N}\beta _{2,1}^{2}}{\beta
_{1,N-1}^{2}}G(x)H_{f}(x)
\end{eqnarray}%
where $H_{f}(x)$ is given by (\ref{Hs}).

The remaining matrix elements $k_{i,j}(x)$ with $i\neq j$ are then%
\begin{equation}
k_{i,j}(x)=\left\{ 
\begin{array}{c}
\displaystyle%
-\frac{\epsilon }{\sqrt{\zeta }}q^{t_{i}-t_{1}}\left( \frac{1-\epsilon q%
\sqrt{\zeta }}{q+1}\right) \frac{\beta _{1,i^{\prime }}\beta _{1,j}}{\beta
_{1,N}}G(x),\quad i<j^{\prime }\quad 2\leq i,j\leq N-1 \\ 
\\ 
\displaystyle%
\frac{1}{\zeta }q^{t_{i}-t_{1}}\left( \frac{1-\epsilon q\sqrt{\zeta }}{q+1}%
\right) \frac{\beta _{1,i^{\prime }}\beta _{1,j}}{\beta _{1,N}}xG(x),\quad
i>j^{\prime }\quad 2\leq i,j\leq N-1%
\end{array}%
\right.
\end{equation}%
and%
\begin{equation*}
k_{2,1}(x)=\frac{2(-1)^{m}(1-\epsilon q\sqrt{\zeta })q^{m-2}}{(1+\epsilon 
\sqrt{\zeta })(\epsilon q^{m-1}\sqrt{\zeta }+(-1)^{m})(\epsilon q^{m}\sqrt{%
\zeta }+(-1)^{m})}\frac{\beta _{1,N-1}}{\beta _{1,N}}G(x),
\end{equation*}%
\begin{equation}
k_{1,m}(x)=-\frac{2\zeta q^{m-1}(q+1)^{2}}{(1-\epsilon q\sqrt{\zeta }%
)(1+\epsilon \sqrt{\zeta })(\epsilon q^{m-1}\sqrt{\zeta }+(-1)^{m})(\epsilon
q^{m}\sqrt{\zeta }+(-1)^{m})}\frac{\beta _{1,N}}{\beta _{1,m^{\prime }}}%
G(x),\quad
\end{equation}%
and the parameters $\beta _{1,j}$ are constrained by the relation 
\begin{equation}
\beta _{1,j}=-\beta _{1,j+1}\frac{\beta _{1,N-j}}{\beta _{1,N+1-j}}%
\;\;\;\;\;\;\;\;\;\;\;\;\;\;\;\;\;j=2,\dots ,m-1.
\end{equation}

With regard to the diagonal matrix elements, they are given by 
\begin{equation}
k_{i,i}(x)\!=\!\left\{ 
\begin{array}{c}
\!k_{1,1}(x)+(\beta _{i,i}-\beta _{1,1})G(x),\quad 
\hfill%
2\leq i\leq m \\ 
\!k_{1,1}(x)+(\beta _{m+1,m+1}-\beta _{1,1})G(x)+\Delta (x),\quad 
\hfill%
i=m+1 \\ 
\!k_{m+1,m+1}(x)+(\beta _{2n+m+1,2n+m+1}-\beta _{m+1,m+1})xG(x)-\epsilon
q^{2n+1}\sqrt{\zeta }\Delta (x),\quad 
\hfill%
i=2n+m+1 \\ 
\!k_{i-1,i-1}(x)+(\beta _{i,i}-\beta _{i-1,i-1})xG(x),\quad 
\hfill%
2n+m+2\leq i\leq N%
\end{array}%
\right.  \label{rr1}
\end{equation}%
The last term of the recurrence relation (\ref{rr1}) is identified with%
\begin{equation}
k_{N,N}(x)=x^{2}k_{1,1}(x)
\end{equation}%
to find%
\begin{equation}
k_{1,1}(x)=[\frac{2x}{x^{2}-1}-\frac{\beta _{m+1,m+1}-\beta _{1,1}}{x+1}%
)]G(x)+[1-\epsilon q^{2n+1}\sqrt{\zeta }]\frac{\Delta (x)}{x^{2}-1}
\end{equation}%
where 
\begin{equation}
\Delta (x)=\frac{2(x-1)G(x)}{(1+\epsilon \sqrt{\zeta })(\epsilon q^{m-1}%
\sqrt{\zeta }+(-1)^{m})(\epsilon q^{m}\sqrt{\zeta }+(-1)^{m})}.
\end{equation}%
In their turn the diagonal parameters $\beta _{i,i}$ are fixed by the
relations 
\begin{equation}
\beta _{i,i}=\left\{ 
\begin{array}{c}
\beta _{1,1}+\Lambda _{m}\dsum\limits_{k=0}^{i-2}(-\frac{1}{q})^{k},\quad 
\hfill%
2\leq i\leq m \\ 
\beta _{1,1}+2-\epsilon q\sqrt{\zeta }\Lambda
_{m}\dsum\limits_{k=0}^{N-1-i}(-q)^{k},\quad 
\hfill%
2n+m+1\leq i\leq N-1 \\ 
\beta _{1,1}+2,\qquad 
\hfill%
i=N%
\end{array}%
\right.
\end{equation}%
and%
\begin{equation}
\beta _{m+1,m+1}=\beta _{1,1}+\Lambda _{m}\left( \frac{q}{q+1}+\frac{(-1)^{m}%
}{q^{m-2}(q+1)^{2}}\frac{1+\epsilon \sqrt{\zeta }}{\epsilon \sqrt{\zeta }}%
\right) ,
\end{equation}%
with 
\begin{equation}
\Lambda _{m}=\frac{2(-1)^{m}q^{m-2}(q+1)^{2}\epsilon \sqrt{\zeta }}{%
(1+\epsilon \sqrt{\zeta })(\epsilon q^{m-1}\sqrt{\zeta }+(-1)^{m})(\epsilon
q^{m}\sqrt{\zeta }+(-1)^{m})}.
\end{equation}%
The class of solution $\mathcal{P}_{3}$ has a total amount of $m$ free
parameters namely $\beta _{1,2n+m+1},\dots ,\beta _{1,N}$.

Here we note that we can take the limit $m=2$ in the class $\mathcal{P}_{3}$
to get a two parameter solution which is a particular case of the three
parameter solution of the class $\mathcal{P}_{2}$. This complete the set of
solutions with the fermionic degree of freedom.

\subsection{K-matrices with bosonic degree of freedom}

Here we consider $k_{i,j}(x)=0,$ for $i\neq j=\{1,...,m\}$ and $i\neq j=$ $%
\left\{ 2n+m+1,...,N\right\} $ in order to get $K$-matrix solutions with
only bosonic degree of freedom. We have found three classes of non-diagonal
solutions that we refer in what follows as solutions of type $\mathcal{P}%
_{4} $ to type $\mathcal{P}_{6}$:

\subsubsection{\textbf{Solution $\mathcal{P}_{4}$:}}

This family of solutions is valid only for the $U_{q}[spo(2|2m)]$ vertex
model with $m\geq 1$ and the corresponding $K$-matrix has the following
block diagonal structure%
\begin{equation}
K^{-}=\left( 
\begin{array}{ccc}
k_{1,1}\mathbb{I}_{m\times m} & \mathbb{O}_{m\times 2} & \mathbb{O}_{m\times
m} \\ 
\mathbb{O}_{2\times m} & 
\begin{array}{cc}
k_{m+1,m+1} & k_{m+1,m+2} \\ 
k_{m+2,m+1} & k_{m+2,m+2}%
\end{array}
& \mathbb{O}_{2\times m} \\ 
\mathbb{O}_{m\times m} & \mathbb{O}_{m\times 2} & k_{N,N}\mathbb{I}_{m\times
m}%
\end{array}%
\right)
\end{equation}%
The non-null entries are given by 
\begin{eqnarray}
k_{1,1}(x) &=&1\;,  \notag \\
k_{m+1,m+1}(x) &=&x\frac{\alpha \lbrack q^{2}\zeta ^{-1}+1](x-1)+2}{\alpha
\lbrack xq^{2}\zeta ^{-1}-1](x-1)+2x},  \notag \\
k_{m+1,m+2}(x) &=&\beta \frac{x(x^{2}-1)}{\alpha \lbrack xq^{2}\zeta
^{-1}-1](x-1)+2x},  \notag \\
k_{m+2,m+1}(x) &=&-\frac{\alpha ^{2}}{\beta }\frac{xq^{2}\zeta ^{-1}(x^{2}-1)%
}{\alpha \lbrack xq^{2}\zeta ^{-1}-1](x-1)+2x},  \notag \\
k_{m+2,m+2}(x) &=&-x^{2}\frac{\alpha \lbrack q^{2}\zeta ^{-1}+1](x-1)-2x}{%
\alpha \lbrack xq^{2}\zeta ^{-1}-1](x-1)+2x},  \notag \\
k_{N,N}(x) &=&x^{2}.
\end{eqnarray}%
where $\alpha =\beta _{m+1,m+1}$ and $\beta =\beta _{m+1,m+2}$ are the two
free parameters.

\subsubsection{\textbf{Solution $\mathcal{P}_{5}$:}}

The solution $\mathcal{P}_{5}$ is admitted for the $U_{q}[spo(4|2m)]$ models
with the following $K$-matrix 
\begin{eqnarray}
&&K^{-}=\left( 
\begin{array}{ccc}
k_{1,1}\mathbb{I}_{m\times m}\! & \mathbb{O}_{m\times 4}\! & \mathbb{O}%
_{m\times m}\! \\ 
\mathbb{O}_{4\times m}\! & 
\begin{array}{cccc}
k_{m+1,m+1} & k_{m+1,m+2} & k_{m+1,m+3} & k_{m+1,m+4} \\ 
k_{m+2,m+1} & k_{m+2,m+2} & k_{m+2,m+3} & k_{m+2,m+4} \\ 
k_{m+3,m+1} & k_{m+3,m+2} & k_{m+3,m+3} & k_{m+3,m+4} \\ 
k_{m+4,m+1} & k_{m+4,m+2} & k_{m+4,m+3} & k_{m+4,m+4}%
\end{array}%
\! & \mathbb{O}_{4\times m}\! \\ 
\mathbb{O}_{m\times m\!}\! & \mathbb{O}_{m\times 4}\! & k_{N,N}\mathbb{I}%
_{m\times m}\!%
\end{array}%
\right) .  \notag \\
&&
\end{eqnarray}%
The non-diagonal elements are all grouped in the $4\times 4$ central block
matrix. With respect to this central block, the entries of the secondary
diagonal are given by 
\begin{eqnarray}
k_{m+1,m+4}(x) &=&\beta _{m+1,m+4}\ G(x)H_{b}(x)  \notag \\
k_{m+2,m+3}(x) &=&\epsilon q^{m-2}\frac{\beta _{m+1,m+3}}{\beta _{m+1,m+2}}%
\Gamma _{m}\ G(x)H_{b}(x),  \notag \\
k_{m+3,m+2}(x) &=&-\epsilon q^{m}\frac{\beta _{m+1,m+2}}{\beta _{m+1,m+3}}%
\Gamma _{m}\ G(x)H_{b}(x),  \notag \\
k_{m+4,m+1}(x) &=&-q^{2m-2}\frac{1}{\beta _{m+1,m+4}}\Gamma _{m}^{2}\
G(x)H_{b}(x),
\end{eqnarray}%
where we recall $%
\displaystyle%
H_{b}(x)=\frac{qx+\epsilon \sqrt{\zeta }}{q+\epsilon \sqrt{\zeta }}$and $%
G(x) $ arbitrary. The remaining non-diagonal elements can be written as 
\begin{eqnarray}
k_{m+1,m+2}(x) &=&\beta _{m+1,m+2}\
G(x)\;\;\;\;\;\;\;\;\;\;\;\;\;\;\;\;\;\;\;\;\;\;\;\;\;k_{m+2,m+1}(x)=-%
\epsilon q^{m-2}\frac{\beta _{m+1,m+3}}{\beta _{m+1,m+4}}\Gamma _{m}G(x), 
\notag \\
k_{m+1,m+3}(x) &=&\beta _{m+1,m+3}\
G(x)\;\;\;\;\;\;\;\;\;\;\;\;\;\;\;\;\;\;\;\;\;\;\;\;\;k_{m+3,m+1}(x)=%
\epsilon q^{m}\frac{\beta _{m+1,m+2}}{\beta _{m+1,m+4}}\Gamma _{m}G(x) 
\notag \\
k_{m+2,m+4}(x) &=&\epsilon q^{m-2}\beta _{m+1,m+3}\
xG(x)\;\;\;\;\;\;\;\;\;\;\;\;\;\;\;k_{m+4,m+2}(x)=q^{2m-2}\frac{\beta
_{m+1,m+2}}{\beta _{m+1,m+4}}\Gamma _{m}\ xG(x)  \notag \\
k_{m+3,m+4}(x) &=&-\epsilon q^{m}\beta _{m+1,m+2}\
xG_{2}(x)\;\;\;\;\;\;\;k_{m+4,m+3}(x)=q^{2m-2}\frac{\beta _{m+1,m+3}}{\beta
_{m+1,m+4}}\Gamma _{m}\ xG(x),  \notag \\
&&
\end{eqnarray}%
where 
\begin{equation}
\Gamma _{m}=\frac{\beta _{m+1,m+2}\beta _{m+1,m+3}}{\beta _{m+1,m+4}}-\frac{2%
}{(q^{m-2}+\epsilon )(q^{m}+\epsilon )}.
\end{equation}%
In their turn the diagonal entries are given by the following expressions 
\begin{eqnarray}
k_{1,1}(x) &=&\left\{ \frac{2(xq^{m-2}+\epsilon )^{2}}{(q^{m-2}+\epsilon
)(q^{m}+\epsilon )(xq^{2m-4}-1)(x-1)}\right.  \notag \\
&&\left. -\frac{\beta _{m+1,m+2}\beta _{m+1,m+3}}{\beta _{m+1,m+4}}\right\} 
\frac{(xq^{m}+\epsilon )(xq^{2m-4}-1)}{(q^{m-2}+\epsilon )x(x+1)}G(x)
\end{eqnarray}%
with the recurrence relation%
\begin{eqnarray}
k_{m+1,m+1}(x) &=&k_{1,1}(x)+(\beta _{m+1,m+1}-\beta _{1,1})G(x)+\Delta
_{1}(x),  \notag \\
k_{m+2,m+2}(x) &=&k_{m+1,m+1}(x)+(\beta _{m+2,m+2}-\beta _{m+1,m+1})G(x), 
\notag \\
k_{m+3,m+3}(x) &=&k_{m+2,m+2}(x)+\Delta _{2}(x),  \notag \\
k_{m+4,m+4}(x) &=&k_{m+3,m+3}(x)+(\beta _{m+4,m+4}-\beta _{m+3,m+3})xG(x), 
\notag \\
k_{N,N}(x) &=&x^{2}k_{1,1}(x),
\end{eqnarray}%
where 
\begin{eqnarray}
\Delta _{1}(x) &=&\frac{\epsilon \Gamma _{m}}{q^{m-2}+\epsilon }\frac{%
(x-1)G(x)}{x}  \notag \\
\Delta _{2}(x) &=&-\epsilon q^{m-2}(q^{2}+1)\Gamma _{m}\left( \frac{%
q^{m-2}x+\epsilon }{q^{m-2}+\epsilon }\right) G(x).
\end{eqnarray}%
The variables $\beta _{i,j}$ are given in terms of the free parameters $%
\beta _{m+1,m+2},\beta _{m+1,m+3}$ and $\beta _{m+1,m+4}$ through the
relations \ here 
\begin{eqnarray}
\beta _{m+1,m+1} &=&\beta _{1,1}-\Gamma _{m}  \notag \\
\beta _{m+2,m+2} &=&\beta _{1,1}+\frac{2}{(q^{m-2}+\epsilon )(q^{m}+\epsilon
)}+\epsilon q^{m-2}\frac{\beta _{m+1,m+2}\beta _{m+1,m+3}}{\beta _{m+1,m+4}}
\notag \\
\beta _{m+3,m+3} &=&\beta _{1,1}+2-\frac{2q^{2m-2}}{(q^{m-2}+\epsilon
)(q^{m}+\epsilon )}-\epsilon q^{m}\frac{\beta _{m+1,m+2}\beta _{m+1,m+3}}{%
\beta _{m+1,m+4}} \\
\beta _{m+4,m+4} &=&\beta _{1,1}+2-\frac{2q^{2m-2}}{(q^{m-2}+\epsilon
)(q^{m}+\epsilon )}+q^{2m-2}\frac{\beta _{m+1,m+2}\beta _{m+1,m+3}}{\beta
_{m+1,m+4}}.
\end{eqnarray}%
We remark here that the form of this class of solution differs from general
form (\ref{Gf}) by the number of free parameters.

\subsubsection{\textbf{Solution $\mathcal{P}_{6}$:}}

The vertex model $U_{q}[spo(2n|2m)]$ admits the solution $\mathcal{P}_{6}$
for $n\geq 3$, whose $K$-matrix has the following block structure 
\begin{equation}
K^{-}=\left( 
\begin{array}{ccc}
k_{1,1}\mathbb{I}_{m\times m} & \mathbb{O}_{m\times 2n} & \mathbb{O}%
_{m\times m} \\ 
\mathbb{O}_{2n\times m} & 
\begin{array}{ccc}
k_{m+1,m+1} & \cdots & k_{m+1,2n+m} \\ 
\vdots & \ddots & \vdots \\ 
k_{2n+m,m+1} & \cdots & k_{2n+m,2n+m}%
\end{array}
& \mathbb{O}_{2n\times m} \\ 
\mathbb{O}_{m\times m} & \mathbb{O}_{m\times 2n} & k_{N,N}\mathbb{I}%
_{m\times m}%
\end{array}%
\right)
\end{equation}

The central block matrix cluster all non-diagonal elements different from
zero. Concerning that central block, we have the following expressions
determining entries of the borders, 
\begin{eqnarray}
k_{i,2n+m}(x) &=&\frac{\epsilon }{\sqrt{\zeta }}\frac{\theta _{i}q^{t_{i}}}{%
\theta _{m+1}q^{t_{m+1}}}\beta _{m+1,i^{\prime
}}xG(x),\;\;\;\;\;\;\;\;\;\;\;\;\;\;\;\;\;\;\;\;\;\;\;\;\;\;\;\;i=m+2,\dots
,2n+m-1  \notag \\
k_{2n+m,j}(x) &=&\frac{\epsilon }{\sqrt{\zeta }}\frac{\theta
_{2n+m}q^{t_{2n+m}}}{\theta _{m+2}q^{t_{m+2}}}\frac{\beta _{m+2,m+1}\beta
_{m+1,j}}{\beta _{m+1,2n+m-1}}xG(x),\;\;\;\;\;\;\;j=m+2,\dots ,2n+m-1  \notag
\\
k_{i,m+1}(x) &=&\frac{\theta _{i}q^{t_{i}}}{\theta _{m+2}q^{t_{m+2}}}\frac{%
\beta _{m+2,m+1}\beta _{m+1,i^{\prime }}}{\beta _{m+1,2n+m-1}}%
G(x),\;\;\;\;\;\;\;\;\;\;\;\;\;\;\;\;\;\;\;\;i=m+3,\dots ,2n+m-1  \notag \\
k_{m+1,j}(x) &=&\beta
_{m+1,j}G(x).\;\;\;\;\;\;\;\;\;\;\;\;\;\;\;\;\;\;\;\;\;\;\;\;\;\;\;\;\;\;\;%
\;\;\;\;\;\;\;\;\;\;\;\;\;\;\;\;\;\;j=m+2,\dots ,2n+m-1
\end{eqnarray}%
The entries of the secondary diagonal are given by 
\begin{equation}
k_{i,i^{\prime }}(x)=\left\{ 
\begin{array}{c}
\displaystyle%
\beta _{m+1,2n+m}G(x)H_{b}(x),\qquad \qquad \qquad \qquad \qquad \qquad
\qquad \qquad \qquad \qquad i=m+1 \\ 
\displaystyle%
-q^{2(m-1)}\frac{\theta _{m+1}q^{t_{m+1}}}{\theta _{i^{\prime
}}q^{t_{i^{\prime }}}}\left( \frac{q+\epsilon \sqrt{\zeta }}{q+1}\right) ^{2}%
\frac{\beta _{m+1,i^{\prime }}^{2}}{\beta _{m+1,2n+m}}G(x)H_{b}(x),\qquad
i=m+2,...,2n+m-1 \\ 
\displaystyle%
\frac{\theta _{2n+m-1}q^{t_{2n+m-1}}}{\theta _{m+2}q^{t_{m+2}}}\frac{\beta
_{m+2,m+1}^{2}\beta _{m+1,2n+m}}{\beta _{m+1,2n+m-1}^{2}}G(x)H_{b}(x),\qquad
\qquad \qquad \qquad \qquad i=2n+m%
\end{array}%
\right.
\end{equation}%
and the remaining non-diagonal elements are determined by the expression 
\begin{equation}
k_{i,j}(x)=\left\{ 
\begin{array}{c}
\displaystyle%
\frac{\epsilon }{\sqrt{\zeta }}\frac{\theta _{i}q^{t_{i}}}{\theta
_{m+1}q^{t_{m+1}}}\left( \frac{q+\epsilon \sqrt{\zeta }}{q+1}\right) \frac{%
\beta _{m+1,i^{\prime }}\beta _{m+1,j}}{\beta _{m+1,2n+m}}G(x),\qquad
i<j^{\prime },\ m+1<i,j<2n+m \\ 
\\ 
\displaystyle%
\frac{1}{\zeta }\frac{\theta _{i}q^{t_{i}}}{\theta _{m+1}q^{t_{m+1}}}\left( 
\frac{q+\epsilon \sqrt{\zeta }}{q+1}\right) \frac{\beta _{m+1,i^{\prime
}}\beta _{m+1,j}}{\beta _{m+1,2n+m}}xG(x),\qquad i>j^{\prime },\ m+1<i,j<2n+m%
\end{array}%
\right. .
\end{equation}%
and%
\begin{eqnarray}
k_{m+1,m+n}(x) &=&-\frac{2\epsilon (-1)^{n}\sqrt{\zeta }(1+q)^{2}}{%
(1-\epsilon \sqrt{\zeta })(q+\epsilon \sqrt{\zeta })}\frac{1}{%
(q^{m}+(-1)^{n}\epsilon )(q^{m-1}-(-1)^{n}\epsilon )}\frac{\beta _{m+1,2n+m}%
}{\beta _{m+1,n+m+1}}G(x),  \notag \\
k_{m+2,m+1}(x) &=&\frac{2\epsilon q(q+\epsilon \sqrt{\zeta })}{\sqrt{\zeta }%
(1-\epsilon \sqrt{\zeta })}\frac{1}{(q^{m}+(-1)^{n}\epsilon
)(q^{m-1}-(-1)^{n}\epsilon )}\frac{\beta _{m+1,2n+m-1}}{\beta _{m+1,2n+m}}%
G(x).
\end{eqnarray}

In their turn the diagonal entries $k_{i,i}(x)$ are given by%
\begin{equation}
k_{1,1}(x)=\left( \frac{x-\epsilon \sqrt{\zeta }}{1-\epsilon \sqrt{\zeta }}%
\right) \left( \frac{xq^{m}+(-1)^{n}\epsilon }{q^{m}+(-1)^{n}\epsilon }%
\right) \left( \frac{xq^{m-1}-(-1)^{n}\epsilon }{q^{m-1}-(-1)^{n}\epsilon }%
\right) \frac{2G(x)}{x(x^{2}-1)}
\end{equation}%
\begin{equation}
k_{i,i}(x)=\left\{ 
\begin{array}{c}
k_{1,1}(x)+\Gamma _{n,m}(x),\quad 
\hfill%
i=m+1 \\ 
k_{m+1,m+1}(x)+(\beta _{i,i}-\beta _{m+1,m+1})G(x),\quad 
\hfill%
i=m+2,...,m+n \\ 
k_{n+m,n+m}(x)+(\beta _{n+m+1,n+m+1}-\beta _{n+m,n+m})xG(x)+\Delta
_{n,m}(x),\quad i=n+m+1 \\ 
k_{n+m+1,n+m+1}(x)+(\beta _{i,i}-\beta _{n+m+1,n+m+1})xG(x),\quad 
\hfill%
i=n+m+2,...,2n+m \\ 
x^{2}k_{1,1}(x),\quad 
\hfill%
i=N%
\end{array}%
\right.
\end{equation}%
where 
\begin{equation}
\Gamma _{n,m}(x)=\frac{2(qx+\epsilon \sqrt{\zeta })}{(1-\epsilon \sqrt{\zeta 
})(q^{m}+(-1)^{n}\epsilon )(q^{m-1}-(-1)^{n}\epsilon )}\frac{G(x)}{x}.
\end{equation}%
and%
\begin{equation*}
\Delta _{n,m}(x)=-\frac{2(-1)^{n}q^{n-1}(q^{2}+1)}{(1-\epsilon \sqrt{\zeta }%
)(q^{m}+(-1)^{n}\epsilon )(q^{m-1}-(-1)^{n}\epsilon )}(x-1)G(x)
\end{equation*}

The diagonal parameters are then fixed by the relations 
\begin{equation}
\beta _{i,i}=\left\{ 
\begin{array}{c}
\displaystyle%
\beta _{1,1}-\frac{q+\epsilon \sqrt{\zeta }}{(1+q)^{2}}Q_{n,m},\qquad 
\hfill%
i=m+1 \\ 
\displaystyle%
\beta _{m+1,m+1}+Q_{n,m}\sum_{k=0}^{i-m-2}(-q)^{k},\qquad 
\hfill%
i=m+2,...,n+m \\ 
\beta _{n+m,n+m}-\epsilon (-1)^{n}(1+q^{2})q^{m-2}\frac{q+\epsilon \sqrt{%
\zeta }}{(1+q)^{2}}Q_{n,m},\qquad 
\hfill%
i=n+m+1 \\ 
\displaystyle%
\beta _{n+m+1,n+m+1}+\epsilon
(-1)^{n}q^{m}Q_{n,m}\sum_{k=0}^{i-n-m-2}(-q)^{k},\qquad 
\hfill%
i=m+2,...,n+m%
\end{array}%
\right.
\end{equation}%
and the auxiliary parameter $Q_{n,m}$ is given by 
\begin{equation}
Q_{n,m}=-\frac{2(1+q)^{2}}{(1-\epsilon \sqrt{\zeta })(q^{m}+(-1)^{n}\epsilon
)(q^{m-1}-(-1)^{n}\epsilon )}.
\end{equation}%
Besides the above relations the following constraints should also holds 
\begin{equation}
\beta _{m+1,m+j}=-\beta _{m+1,j+m+1}\frac{\beta _{m+1,2n+m-j}}{\beta
_{m+1,2n+m+1-j}}\;\;\;\;\;\;\;\;\;\;\;\;\;\;\;\;\;\;\;\;\;\;j=2,\dots ,n-1,
\end{equation}%
and $\beta _{m+1,m+n+1},\dots ,\beta _{m+1,2n+m}$ are regarded as the $n$
free parameters. The case $n=2$ is a particular solution of the three
parameter class $\mathcal{P}_{6}$.

\subsection{Complete K-matrices}

The complete $K$-matrices are solutions with all entries different from
zero. This kind of solution will be present only in two class: \ the models
with two bosonic degree of freedom, $U_{q}[spo(2|2m)]$ \ and those models
with only two fermionic degree of freedom $U_{q}[spo(2n|2)]$ $.$

\subsubsection{\textbf{Solution $\mathcal{P}_{7}$:}}

The series of solutions $\mathcal{P}_{7}$ is valid for the $U_{q}[spo(2|2m)]$
model and the corresponding $K$-matrix also possess all entries different
from zero. In the first and last columns, the matrix elements are mainly
given by 
\begin{eqnarray}
k_{i,1}(x) &=&\frac{\theta _{i}q^{t_{i}}}{\theta _{2}q^{t_{2}}}\frac{\beta
_{2,1}\beta _{1,i^{\prime }}}{\beta _{1,N-1}}G(x)\;\;\;\;\;\;\;\;\;\;\;\;\;%
\;\;\;\;\;\;i=3,\dots ,N-1  \notag \\
k_{i,N}(x) &=&-\frac{\epsilon _{m}}{\sqrt{\zeta }}\frac{\theta _{i}q^{t_{i}}%
}{\theta _{1}q^{tt_{1}}}\beta _{1,i^{\prime
}}xG(x)\;\;\;\;\;\;\;\;\;\;\;\;\;\;\;\;i=2,\dots ,N-1
\end{eqnarray}%
while the ones in the first and last rows are respectively 
\begin{eqnarray}
k_{1,j}(x) &=&\beta
_{1,j}G(x)\;\;\;\;\;\;\;\;\;\;\;\;\;\;\;\;\;\;\;\;\;\;\;\;\;\;\;\;\;\;\;\;\;%
\;\;\;\;\;\;\;\;\;\;\;j=2,\dots ,N-1  \notag \\
k_{N,j}(x) &=&-\frac{\epsilon _{m}}{\sqrt{\zeta }}\frac{\theta _{N}q^{t_{N}}%
}{\theta _{2}q^{t_{2}}}\frac{\beta _{2,1}\beta _{1,j}}{\beta _{1,N-1}}%
xG(x)\;\;\;\;\;\;\;\;\;\;\;\;j=2,\dots ,N-1
\end{eqnarray}

In the secondary diagonal we have the following expression determining the
matrix elements 
\begin{equation}
k_{i,i^{\prime }}(x)=\left\{ 
\begin{array}{c}
\beta _{1,N}G(x)H_{f}(x),\qquad 
\hfill%
i=1 \\ 
\displaystyle%
\frac{\theta _{1}q^{t_{1}}}{\theta _{i^{\prime }}q^{t_{i^{\prime }}}}\left( 
\frac{1\epsilon _{m}\epsilon q\sqrt{\zeta }}{q+1}\right) ^{2}\frac{\beta
_{1,i^{\prime }}^{2}}{\beta _{1,N}}G(x)H_{f}(x),\qquad 
\hfill%
i\neq \{1,m+1,m+2,N\} \\ 
\displaystyle%
\frac{\theta _{1}q^{t_{1}}}{\theta _{i^{\prime }}q^{t_{i^{\prime }}}}\left( 
\frac{q\epsilon _{m}\epsilon \sqrt{\zeta }}{q-1}\right) \left( \frac{%
1-\epsilon _{m}q\sqrt{\zeta }}{q+1}\right) \frac{\beta _{1,i^{\prime }}^{2}}{%
\beta _{1,N}}G(x)H_{b}(x),\qquad i=\{m+1,m+2\} \\ 
\displaystyle%
\frac{\theta _{N-1}q^{t_{N-1}}}{\theta _{2}q^{t_{2}}}\frac{\beta _{1,N}\beta
_{2,1}^{2}}{\beta _{1,N-1}^{2}}G(x)H_{f}(x),\qquad 
\hfill%
i=N.%
\end{array}%
\right.
\end{equation}%
recalling that 
\begin{equation}
H_{b}(x)=\frac{qx+\epsilon _{m}\sqrt{\zeta }}{q+\epsilon _{m}\sqrt{\zeta }}%
\;\;\;\mathrm{and}\;\;\;H_{f}(x)=\frac{x-\epsilon _{m}q\sqrt{\zeta }}{%
1-\epsilon _{m}q\sqrt{\zeta }}.  \label{hfhb}
\end{equation}

In their turn the other non-diagonal entries satisfy the relation 
\begin{equation}
k_{i,j}(x)=\left\{ 
\begin{array}{c}
\displaystyle%
-\frac{\epsilon _{m}}{\sqrt{\zeta }}\frac{\theta _{i}q^{t_{i}}}{\theta
_{1}q^{t_{1}}}\left( \frac{1-\epsilon _{m}q\sqrt{\zeta }}{q+1}\right) \frac{%
\beta _{1,i^{\prime }}\beta _{1,j}}{\beta _{1,N}}G(x),\qquad i<j^{\prime },\
2<i,j<N-1 \\ 
\displaystyle%
\frac{1}{\zeta }\frac{\theta _{i}q^{t_{i}}}{\theta _{1}q^{t_{1}}}\left( 
\frac{1-\epsilon _{m}q\sqrt{\zeta }}{q+1}\right) \frac{\beta _{1,i^{\prime
}}\beta _{1,j}}{\beta _{1,N}}xG(x),\qquad i>j^{\prime },\ 2<i,j<N-1%
\end{array}%
\right. ,
\end{equation}%
and%
\begin{eqnarray}
k_{1,m}(x) &=&\frac{i(q+1)}{q-1}\frac{\beta _{1,m+1}\beta _{1,m+2}}{\beta
_{1,m+3}}G(x),  \notag \\
k_{1,m+1}(x) &=&%
\displaystyle%
(-1)^{\delta _{m,2}}\frac{i\sqrt{\zeta }(q+\epsilon _{m})}{(1+\epsilon _{m}%
\sqrt{\zeta })(1+q\sqrt{\zeta })}\frac{\beta _{1,N}}{\beta _{1,m+2}}G(x), 
\notag \\
k_{2,1}(x) &=&(-1)^{\delta _{m,2}}\frac{\epsilon _{m}}{q\sqrt{\zeta }}\frac{%
(1-\epsilon _{m}q\sqrt{\zeta })^{2}}{(q-\epsilon _{m})(1+\epsilon _{m}\sqrt{%
\zeta })(1+q\sqrt{\zeta })}\frac{\beta _{1,N-1}}{\beta _{1,N}}G(x),
\label{m2}
\end{eqnarray}%
and the parameters $\beta _{1,j}$ are required to satisfy the recurrence
relation 
\begin{equation}
\beta _{1,j}=-\frac{\beta _{1,j+1}\beta _{1,N-j}}{\beta _{1,N-j+1}}%
\;\;\;\;\;\;\;\;\;\;\;\;\;\;\;j=2,\dots ,m-1.
\end{equation}

Considering now the diagonal entries, they are given by%
\begin{equation}
k_{i,i}(x)=\left\{ 
\begin{array}{c}
k_{1,1}+(\beta _{i,i}-\beta _{1,1})G(x),\qquad 
\hfill%
i=2,...,m+1 \\ 
k_{1,1}(x)+(\beta _{m+2,m+2}-\beta _{1,1})xG(x)+\Delta (x),\qquad 
\hfill%
i=m+2 \\ 
k_{m+2,m+2}(x)+(\beta _{i,i}-\beta _{m+2,m+2})xG(x),\qquad 
\hfill%
i=m+3,...,N-1 \\ 
x^{2}k_{1,1}(x).\quad 
\hfill%
i=N%
\end{array}%
\right. ,
\end{equation}%
where%
\begin{equation*}
k_{1,1}(x)=\frac{G(x)}{x-1}\qquad \mathrm{and}\qquad \Delta (x)=(1-x)G(x).
\end{equation*}%
The diagonal parameters are determined by the expressions 
\begin{equation}
\beta _{i,i}=\left\{ 
\begin{array}{c}
\displaystyle%
\beta _{1,1}+\frac{(q+1)}{q(1+\epsilon _{m}\sqrt{\zeta })}\sum_{k=0}^{i-2}(-%
\frac{1}{q})^{k},\qquad 
\hfill%
i=2,...,m \\ 
\displaystyle%
\beta _{m,m}-\frac{2\epsilon _{m}\sqrt{\zeta }}{q(q+1)(1+\epsilon _{m}\sqrt{%
\zeta })},\qquad 
\hfill%
i=m+1 \\ 
\displaystyle%
\beta _{m+1,m+1}+\frac{(q^{2}+1)(q+\epsilon _{m}\sqrt{\zeta })}{%
q(q+1)(1+\epsilon _{m}\sqrt{\zeta })},\qquad 
\hfill%
i=m+2 \\ 
\displaystyle%
\beta _{m+2,m+2}-\frac{2q^{2}}{(q+1)(1+\epsilon _{m}\sqrt{\zeta })},\qquad 
\hfill%
i=m+3 \\ 
\displaystyle%
\beta _{m+3,m+3}+\left( \frac{q+1}{1+\epsilon _{m}\sqrt{\zeta }}\right)
\sum_{k=0}^{i-m-4}(-\frac{1}{q})^{k}.\qquad 
\hfill%
i=m+4,...,N%
\end{array}%
\right.
\end{equation}%
This solution has altogether $m+1$ free parameters, namely $\beta
_{1,m+2},\dots ,\beta _{1,N}$. This complete solution depend on the parity
of $m$ through the relation $\epsilon _{m}=(-1)^{m}.$

Here we notice that the case $m=2$ is also special because $\beta
_{1,m}=\beta _{1,2}$ and we have an exchange of sign for $k_{2,1}(x)$ and $%
k_{1,3}(x)$ as indicate in equation (\ref{m2}).

\subsubsection{\textbf{Solution $\mathcal{P}_{8}$:}}

The class of solutions $\mathcal{P}_{8}$ is valid for the vertex model $%
U_{q}[spo(2n|2)]$ and the corresponding $K$-matrix contains only non-null
entries. The border elements are mainly given by the following expressions 
\begin{eqnarray}
k_{i,N}(x) &=&-\frac{\epsilon _{n}}{\sqrt{\zeta }}\frac{\theta _{i}q^{t_{i}}%
}{\theta _{1}q^{t_{1}}}\beta _{1,i^{\prime
}}xG(x)\;\;\;\;\;\;\;\;\;\;\;\;\;\;\;\;\;\;\;\;\;i=2,\dots ,N-1  \notag \\
k_{i,1}(x) &=&\frac{\theta _{i}q^{t_{i}}}{\theta _{2}q^{t_{2}}}\frac{\beta
_{2,1}\beta _{1,i^{\prime }}}{\beta _{1,N-1}}G(x)\;\;\;\;\;\;\;\;\;\;\;\;\;%
\;\;\;\;\;\;\;\;\;\;\;\;i=3,\dots ,N-1  \notag \\
k_{N,j}(x) &=&-\frac{\epsilon _{n}}{\sqrt{\zeta }}\frac{\theta _{N}q^{t_{N}}%
}{\theta _{2}q^{t_{2}}}\frac{\beta _{2,1}\beta _{1,j}}{\beta _{1,N-1}}%
xG(x)\;\;\;\;\;\;\;\;\;\;\;\;\;\;j=2,\dots ,N-1  \notag \\
k_{1,j}(x) &=&\beta
_{1,j}G(x)\;\;\;\;\;\;\;\;\;\;\;\;\;\;\;\;\;\;\;\;\;\;\;\;\;\;\;\;\;\;\;\;\;%
\;\;\;\;\;\;\;\;\;\;\;\;\;j=2,\dots ,N-1.
\end{eqnarray}

The secondary diagonal is constituted by elements $k_{i,i^{\prime }}(x)$
given by%
\begin{equation}
k_{i,i^{\prime }}(x)=\left\{ 
\begin{array}{c}
\beta _{1,N}G(x)H_{f}(x),\qquad 
\hfill%
i=1 \\ 
\displaystyle%
-\frac{\theta _{1}q^{t_{1}}}{\theta _{i^{\prime }}q^{t_{i^{\prime }}}}\left( 
\frac{1-\epsilon _{n}q\sqrt{\zeta }}{q-1}\right) \left( \frac{q+\epsilon _{n}%
\sqrt{\zeta }}{q+1}\right) \frac{\beta _{1,i^{\prime }}^{2}}{\beta _{1,N}}%
G(x)H_{b}(x),\qquad i\neq \{1,N\} \\ 
\displaystyle%
\frac{\theta _{N-1}q^{t_{N-1}}}{\theta _{2}q^{t_{2}}}\frac{\beta _{1,N}\beta
_{2,1}^{2}}{\beta _{1,N-1}^{2}}G(x)H_{f}(x),\qquad 
\hfill%
i=N.%
\end{array}%
\right.
\end{equation}%
where the functions $H_{b}(x)$ and $H_{f}(x)$ were already given in (\ref%
{hfhb}). The remaining non-diagonal entries are determined by the expression 
\begin{equation}
k_{i,j}(x)=\left\{ 
\begin{array}{c}
\displaystyle%
\frac{\epsilon _{n}}{\sqrt{\zeta }}\frac{\theta _{i}q^{t_{i}}}{\theta
_{1}q^{t_{1}}}\left( \frac{1-\epsilon _{n}q\sqrt{\zeta }}{q-1}\right) \frac{%
\beta _{1,i^{\prime }}\beta _{1,j}}{\beta _{1,N}}G(x),\qquad i<j^{\prime },\
2<i,j<N-1 \\ 
\displaystyle%
\frac{1}{\zeta }\frac{\theta _{i}q^{t_{i}}}{\theta _{1}q^{t_{1}}}\left( 
\frac{1-\epsilon _{n}q\sqrt{\zeta }}{q-1}\right) \frac{\beta _{1,i^{\prime
}}\beta _{1,j}}{\beta _{1,N}}xG(x),\qquad i>j^{\prime },\ 2<i,j<N-1%
\end{array}%
\right. ,
\end{equation}%
and%
\begin{equation*}
k_{1,n+1}(x)=%
\displaystyle%
\frac{-i\sqrt{\zeta }(q+1)}{(q\sqrt{\zeta }-\epsilon _{n})(\sqrt{\zeta }%
-\epsilon _{n})}\frac{\beta _{1,N}}{\beta _{1,n+2}}G(x)
\end{equation*}%
\begin{equation*}
k_{2,1}(x)=\frac{-i\epsilon _{n}(q\sqrt{\zeta }-\epsilon _{n})}{(\sqrt{\zeta 
}-\epsilon _{n})(q-1)}\frac{\beta _{1,N-1}}{\beta _{1,N}}G(x).
\end{equation*}%
and the parameters $\beta _{1,j}$ are required to satisfy 
\begin{equation}
\beta _{1,j}=-\frac{\beta _{1,j+1}\beta _{1,N-j}}{\beta _{1,N+1-j}}%
\;\;\;\;\;\;\;\;\;\;\;\;\;\;\;\;\;\;\;\;\;j=2,\dots ,n.
\end{equation}

With respect to the diagonal entries, they are given by%
\begin{equation}
k_{i,i}(x)=\left\{ 
\begin{array}{c}
k_{1,1}(x)+(\beta _{i,i}-\beta _{1,1})G(x),\qquad 
\hfill%
i=2,...,n+1 \\ 
k_{n+1,n+1}(x)+(\beta _{n+2,n+2}-\beta _{n+1,n+1})xG(x)+\Delta (x),\qquad 
\hfill%
i=n+2 \\ 
k_{n+2,n+2}(x)+(\beta _{i,i}-\beta _{n+2,n+2})xG(x),\qquad 
\hfill%
i=n+3,...,N-1 \\ 
x^{2}k_{1,1}(x).\qquad 
\hfill%
i=N%
\end{array}%
\right. ,
\end{equation}%
where%
\begin{equation*}
k_{1,1}(x)=\frac{G(x)}{x-1}\qquad \mathrm{and}\qquad \Delta (x)=\epsilon
_{n}\left( \frac{\sqrt{\zeta }+q}{\sqrt{\zeta }+\epsilon _{n}q}\right)
\left( \frac{q^{2}+\epsilon _{n}}{\sqrt{\zeta }-\epsilon _{n}}\right) \left( 
\frac{x-1}{q-1}\right) G(x).
\end{equation*}%
The parameters $\beta _{i,i}$ are determined by the expressions 
\begin{equation}
\beta _{i,i}=\left\{ 
\begin{array}{c}
\displaystyle%
\beta _{1,1}+\frac{(-1)^{n+i}[q^{i-1}+q^{i-2}-(-1)^{i}(q-1)]}{(q-1)(\sqrt{%
\zeta }-\epsilon _{n})},\qquad 
\hfill%
i=2,...,n+1 \\ 
\displaystyle%
\beta _{n+1,n+1}+\frac{(\sqrt{\zeta }+q)(q^{2}+\epsilon _{n})}{q(q-1)(\sqrt{%
\zeta }-\epsilon _{n})},\qquad 
\hfill%
i=n+2 \\ 
\displaystyle%
\beta _{n+2,n+2}-\frac{\epsilon _{n}q(q+1)^{2}}{(q-1)(\sqrt{\zeta }-\epsilon
_{n})}\sum_{k=0}^{i-n-3}(-q)^{k},\qquad 
\hfill%
i=n+3,...,N-1 \\ 
\beta _{1,1}+2.\qquad 
\hfill%
i=N%
\end{array}%
\right.
\end{equation}%
The variables $\beta _{1,n+2},\dots ,\beta _{1,N}$ give us a total amount of 
$n+1$ free parameters. This complete solution depend on the parity of $n$
through the relation $\epsilon _{n}=(-1)^{n}.$

\section{Concluding Remarks}

In this work we have presented the general set of regular solutions of the
graded reflection equation for the $U_{q}[osp(2n|2m)]$ vertex model. Our
findings can be summarized into four classes of diagonal solutions and eight
classes of non-diagonal ones. Although the $R$ matrix of the $%
U_{q}[osp(2m|2n)^{(1)}]$ vertex model is isomorphic to the $R$ matrix of the 
$U_{q}[spo(2n|2m)]$ vertex model, we have find additional different $K$%
-matrices from those obtained by the exchange of the degree of freedom in
these models (see, for example, the diagonal solutions of the $%
U_{q}[spo(2|2)]$ vertex model).

We expect the results presented here to motivate further developments on the
subject of integrable open boundaries for vertex models based on $q$%
-deformed Lie superalgebras. In particular, the classification of the
solutions of the graded reflection equation for others $q$-deformed Lie
superalgebras, for instance, the $U_{q}[osp(2n|2m)^{\left( 2\right) }]$ Lie
superalgebra, which we hope to report on a future work.

\section{Acknowledgments}

We thank to W. Galleas and M.J. Martins by the discussions. This work is
partially supported by the Brazilian research councils CNPq and FAPESP.

\section*{\textbf{Appendix A: The $U_{q}[spo(2|2)]$ case}}

\setcounter{equation}{0} \renewcommand{\theequation}{B.\arabic{equation}}
The set of $K$-matrices associated with the $U_{q}[spo(2|2)]$ vertex model
includes both diagonal and non-diagonal solutions. The four solutions
intrinsically diagonal three contain only one free parameter $\beta $ and
they are given by 
\begin{eqnarray}
K^{-}(x) &=&\mathrm{Diag}(\frac{2x-\beta (x-1)}{2x+\beta x(x-1)},1,1,\frac{%
2x+\beta x(xq^{4}-1)}{2x+\beta (q^{4}-x)}),  \notag \\
K^{-}(x) &=&\mathrm{Diag}(1,1,\frac{2x+\beta x(x-1)}{2x-\beta (x-1)},\frac{%
2x+\beta x(x-1)}{2x-\beta (x-1)})  \notag \\
K^{-}(x) &=&\mathrm{Diag}(1,\frac{2x+\beta x(x-1)}{2x-\beta (x-1)},x^{2},%
\frac{2x+\beta x(x-1)}{2x-\beta (x-1)}x^{2})
\end{eqnarray}%
and one without free parameters 
\begin{equation}
K^{-}(x)=\mathrm{Diag}(1,x\frac{q^{2}+\epsilon x}{xq^{2}+\epsilon },x\frac{%
q^{2}+\epsilon x}{xq^{2}+\epsilon },x^{2}).
\end{equation}%
We have also found the following non-diagonal solutions 
\begin{equation}
K^{-}(x)=\left( 
\begin{array}{cccc}
1\!-\!\frac{(x^{-1}\beta ^{2}+\alpha )(x-1)}{2\beta } & 0 & 0 & \frac{1}{2}%
\!\beta _{23}(x^{2}-1) \\ 
0 & 1\!+\!\frac{(\beta ^{2}-\alpha )(x-1)}{2\beta } & 0 & 0 \\ 
0 & 0 & x^{2}\!-\!\frac{(\beta ^{2}-\alpha )x(x-1)}{2\beta } & 0 \\ 
\frac{\alpha }{2\beta _{23}}(x^{2}-1) & 0 & 0 & x^{2}\!+\frac{(x\beta
^{2}+\alpha )x(x-1)}{2\beta }%
\end{array}%
\right)
\end{equation}%
containing three free parameters $\alpha ,\beta $ and $\beta _{23}$, and
another solution with one free parameter 
\begin{equation}
K^{-}(x)=\left( 
\begin{array}{cccc}
1 & 0 & 0 & \frac{1}{2}\beta _{1,4}(x^{2}-1) \\ 
0 & \frac{q^{2}-x^{2}}{q^{2}-1} & 0 & 0 \\ 
0 & 0 & \frac{q^{2}-x^{2}}{q^{2}-1} & 0 \\ 
0 & \frac{2}{\beta _{1,4}}\frac{q^{2}}{(q^{2}-1)^{2}}(x^{2}-1) & 0 & x^{2}%
\end{array}%
\right) .  \label{b5}
\end{equation}%
We don't find any complete solution for this model.

\end{document}